# An Alternative Framework for Time Series Decomposition and Forecasting and its Relevance for Portfolio Choice – A Comparative Study of the Indian Consumer Durable and Small Cap Sectors

Jaydip Sen
Calcutta Business School, Diamond Harbor Road, Bishnupur – 743503
West Bengal, INDIA
email: jaydip.sen@acm.org

and

Tamal Datta Chaudhuri
Calcutta Business School, Diamond Harbour Road, Bishnupur – 743503
West Bengal, INDIA
email: tamalc@calcuttabusinessschool.org

**ABSTRACT**

One of the challenging research problems in the domain of time series analysis and forecasting is making efficient and robust prediction of stock market prices. With rapid development and evolution of sophisticated algorithms and with the availability of extremely fast computing platforms, it has now become possible to effectively extract, store, process and analyze high volume stock market time series data. Complex algorithms for forecasting are now available for speedy execution over parallel architecture leading to fairly accurate results. In this paper, we have used time series data of the two sectors of the Indian economy – Consumer Durables sector and the Small Cap sector for the period January 2010 – December 2015 and proposed a decomposition approach for better understanding of the behavior of each of the time series. Our contention is that various sectors reveal different time series patterns and understanding them is essential for portfolio formation. Further, based on this structural analysis, we have also proposed several robust forecasting techniques and analyzed their accuracy in prediction using suitably chosen training and test data sets. Extensive results are presented to demonstrate the effectiveness of our propositions.

**Key words:** Time Series, Decomposition, Trend, Seasonal, Random, Holt Winters Forecasting model, ARIMA, R Programming Language, BSE Consumer Durables Index, BSE Small Cap Index.

**JEL Classification:** G 11, G 14, G 17, C 63

## 1. Introduction

Prediction of stock prices has been one of the biggest challenges to researchers, particularly to those belonging to the Artificial Intelligence (AI) community. Various technical, fundamental, and statistical indicators have been proposed and used with varying results. In our recent research work, we have proposed a new way of looking at portfolio diversification and prediction of stock returns (Sen & Datta Chaudhuri, 2016a; Sen & Datta Chaudhuri, 2016b). It has been postulated that different sectors in an economy do not behave uniformly, and sectors differ from each other in terms of their trend pattern, their seasonal characteristics and also their randomness. While the randomness aspect has been the cornerstone of *Efficient Market Hypothesis*, the literature trying to prove or disprove it, has delved into the various fundamental characteristics of each company and have come up with different results. For example, Datta Chaudhuri, Ghosh and Eram applied Random Forest and Dynamic Evolving Neural-Fuzzy Inference System (DENFIS) to predict stock returns of mid cap Indian firms (Datta Chaudhuri et al, 2016). Our contention has been that, besides their fundamental characteristics, performances of companies depend on the performance of the sector to which they belong, and each sector has its own reason for growth or stagnation. The reasons behind the fortunes of the IT sector in India is different from those of the Steel sector or the Pharmaceutical sector, and these differences have to be factored in for portfolio choice and also churning of the portfolio.

In this work, we focus on the time series pattern of two sectors in India, namely the Consumer Durables sector and the Small Cap sector. We first demonstrate that the time series decomposition approach proposed provides us with deeper understanding of the behavior of a time series by observing the relative magnitudes of its three components namely trend, seasonal and random and also enables us to validate some hypotheses. For example, the Consumer Durables sector in India is known to display seasonal characteristics and the Small Cap sector in India is speculative in nature, and hence should have strong random components. The decomposition approach enables us to study the seasonal components and the random components of these two sectors separately and validate these hypotheses. With regard to the seasonal components, the decomposition approach also helps us to understand during which months which sectors are strong/weak so that buy/sell decisions about the stocks of companies in those sectors can be made effectively. The sectors with dominant random components in their time series, however, can be used for pure speculative gains.

Second, we propose an extensive framework for time series forecasting and a quantitative approach to analyze the change in behavior of the constituents (i.e., the trend, the seasonal, and the random component) of a time series over a long period of time. We have applied five techniques of forecasting using R environment and also provided a detailed guideline about which technique to use under what situations and for what type of time series behavior.

Third, we have presented a robust quantitative approach for analyzing any change in behavior of the constituents (i.e., the trend and the seasonal component) of a time series over a long period of time. If the behavior of the components of a time series does not change significantly over time, it is possible to design very robust forecasting framework for the time series.

The rest of the paper is organized as follows. Section 2 briefly discusses the methodology in constructing various time series and decomposing the time series into its components. It also presents a brief outline on the forecasting frameworks designed in this work using the R programming language. Section 3 provides a detailed discussion on the methods of decomposition, the decomposition results of both sectors under study, and an analysis of the results. In addition, it presents two hypotheses and their validations using our experimental results. In Section 4, five robust forecasting techniques are proposed and a framework for analyzing the behavior of the structural constituents (i.e., the trend, the seasonal, and the random component) of a time series using the R programming environment. Section 5 presents detailed results of forecasting using all the methods that we proposed in Section 4. The forecasting methods are compared based on some suitably chosen metrics and a critical comparative analysis is presented for the proposed methods of forecasting. We have also analyzed the reason why certain methods perform better compared to the others for certain time series under certain situations. In Section 6, we discuss some related work in the current literature. Finally, Section 7 concludes the paper.

## 2. Methodology

In this section, we provide a brief outline of the methodology that we have followed in our work. However, each of the following sections contain detailed discussion on the methodology followed in the work related to that Section. We have used the *R programming language* (Ihaka & Gentleman, 1996) for data management, data analysis and presentation of results. R is an open source language with very rich libraries that is ideally suited for data analysis work. In this work, we use daily data from the Bombay Stock Exchange (BSE) on BSE Consumer Durables Index and BSE Small Cap index for the period January 2010 to December 2015. The daily index values are first stored in two plain text files – each sector data in one file. The daily data are then aggregated into monthly averages resulting in 70 values in the time series data. These 70 monthly average values for each sector are stored into two different plain text files – each sector monthly average in one file. The records in the text file for each sector are read into an R variable using the *scan( )* function in R. The resultant R variable is converted into a monthly time series variable using the *ts( )* function defined in the *TTR* library in the *R* programming language. The monthly time series variable in R is now an aggregate of its three constituent components: (i) trend, (ii) seasonal, and (iii) random. We then decompose the time series into its three components. For this purpose, we use the *decompose( )* function defined in the *TTR* library in R. The decomposition results enable us to make a comparative analysis of the behavior of the two different time series belonging to two different sectors. We validate two hypotheses by our deeper analysis of the decomposition results.

After a detailed analysis of the decomposition results, we enter into our second endeavor in this work. We have designed and analyzed five robust forecasting methods using the *HoltWinters( )* function, *Auto Regressive Integrated Moving Average* (ARIMA) framework, and an approach based on computation of the aggregate of the trend and seasonal components – all in the *R* computing framework. A detailed comparative analysis, highlighting which method performs best under what situation and for what type of time series, is also presented.

In our previous work, we have highlighted the effectiveness of time series decomposition approach for robust analysis and forecasting of the Indian Auto sector (Sen & Datta Chaudhuri, 2016a; Sen & Datta Chaudhuri, 2016b). In this work, we have compared two different sectors – Indian Consumer Durables sector and the Indian Small Cap sector and proposed guidelines and frameworks for comparing different sectors based on time series decomposition studies. Based on our analysis, we have also validated two hypotheses on the behavior of the two sectors under study. We have also analyzed and determined what forecasting technique to use based on the behavior of the time series and also have highlighted the reasons why some forecasting approaches perform better in comparison with other approaches under certain situations.

## 3. Time Series Decomposition Results

We now present the methods that we have followed to decompose time series for both BSE Consumer Durables Index and BSE Small Cap Index and then present the results that we have obtained from the decomposition work.

For both the sectors, we have first taken the daily index values from January 2010 to December 2015 and saved the values in plain text (.txt) files. From these daily index values, we have computed the month averages and saved the monthly average values in two different text files. Each of these text files contained 72 values (6 years, each year containing 12 month average values). We used R language function *scan( )* to read these text files and store them into appropriate R variables. Then, we converted these R variables into time series variables using the R function *ts( )* defined in the package *TTR*. Once these time series variables are constructed, we have used the *plot( )* function in R to derive the displays of the time series. The time series for the Consumer Durables sector and the Small Cap sector are presented in Figure 1 and Figure 3 respectively.

The plots of the time series for the two sectors provide us an overall idea about how the two sectors have performed over the period under consideration (i.e., January 2010 – December 2015). Figure 2 and Figure 4 present the results of decomposition for the times series of the Consumer Durables sector and the Small Cap sector respectively. Each of these two figures have four boxes arranged in a stack. The boxes depict the overall time series, the trend, the seasonal and the random component respectively, arranged from top to bottom.

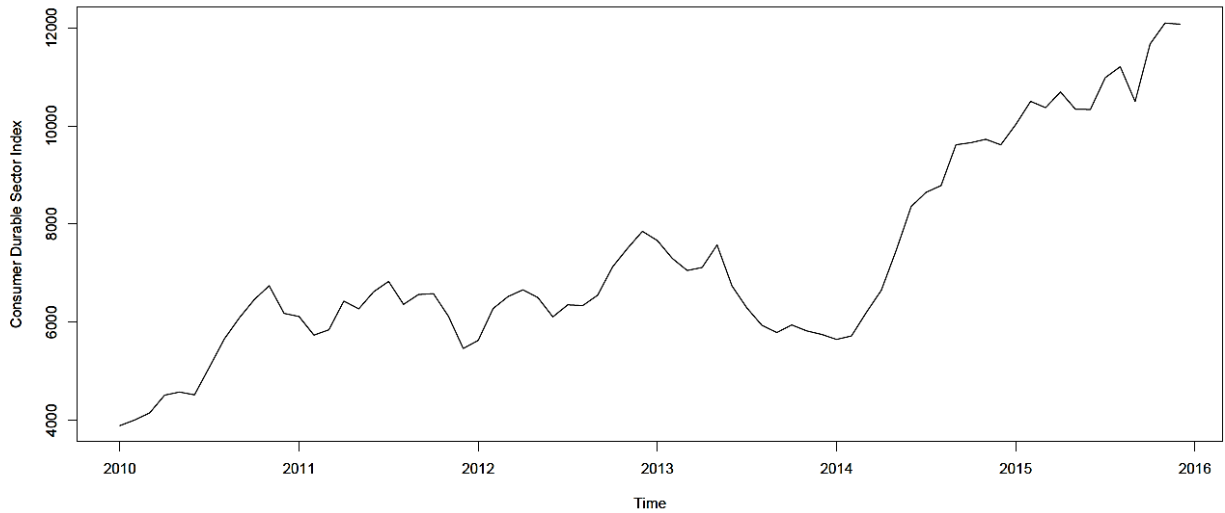

**Figure 1: The consumer durables sector index time series (Jan 2010 – Dec 2015)**

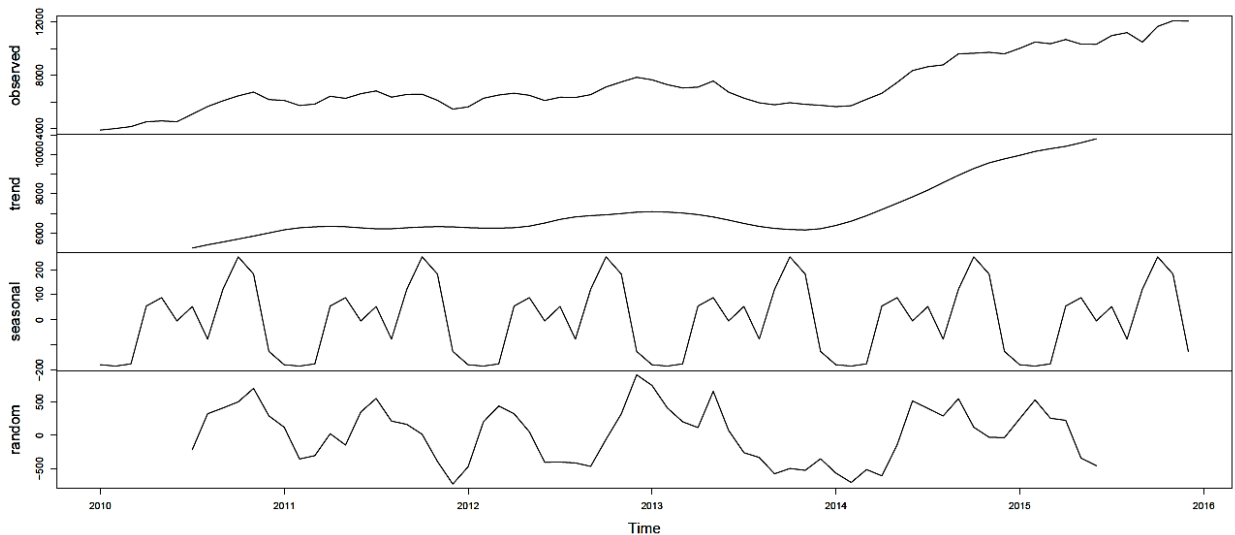

**Figure 2: Decomposition of consumer durables sector index time series into its trend, seasonal and random components (Jan 2010 – Dec 2015)**

Table 1 and Table 2 present the numerical values of the time series data and its three components for the Consumer Durables sector and the Small Cap sector respectively. It may be interesting to observe that the values of the trend and the random components are not available for the period January 2010 – June 2010 and also for the period July 2015 – December 2015. Since the *decompose( )* function in R uses a 12 month moving average method for computing the trend component, in order to compute the trend value for January 2010, we need time series data from July 2009 to June 2010. However, since we have used time series data from January 2010 to December 2015, the first trend value the *decompose( )* function could compute was for the month of July 2010 and the last month being June 2014. For computing the seasonal component, the

*decompose( )* function first *detrends* (subtracts the trend component from the overall time series) the time series and arranges the time series values in a 12 column format. The seasonal values for each month is derived by computing the averages of each column. The value of the seasonal component for a given month remains the same for the entire period under study. The random components are obtained after subtracting the sum of the corresponding trend and seasonal components from the overall time series values. Since the trend values for the period January 2010 – June 2010 and July 2015 – December 2015 are missing, the random components for those periods could not be computed as well.

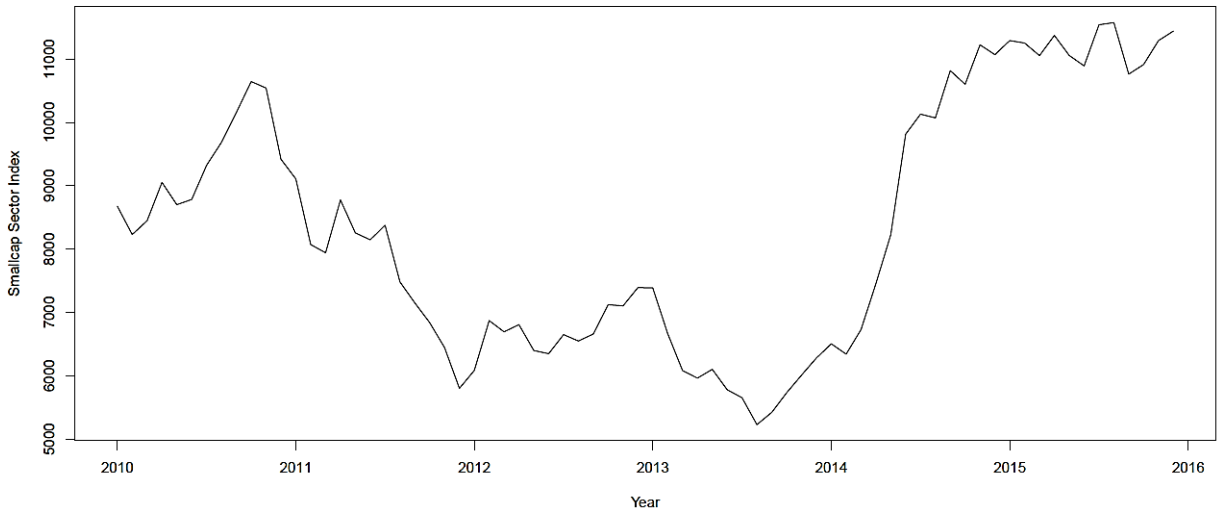

**Figure 3: The small cap sector index time series (Jan 2010 – Dec 2015)**

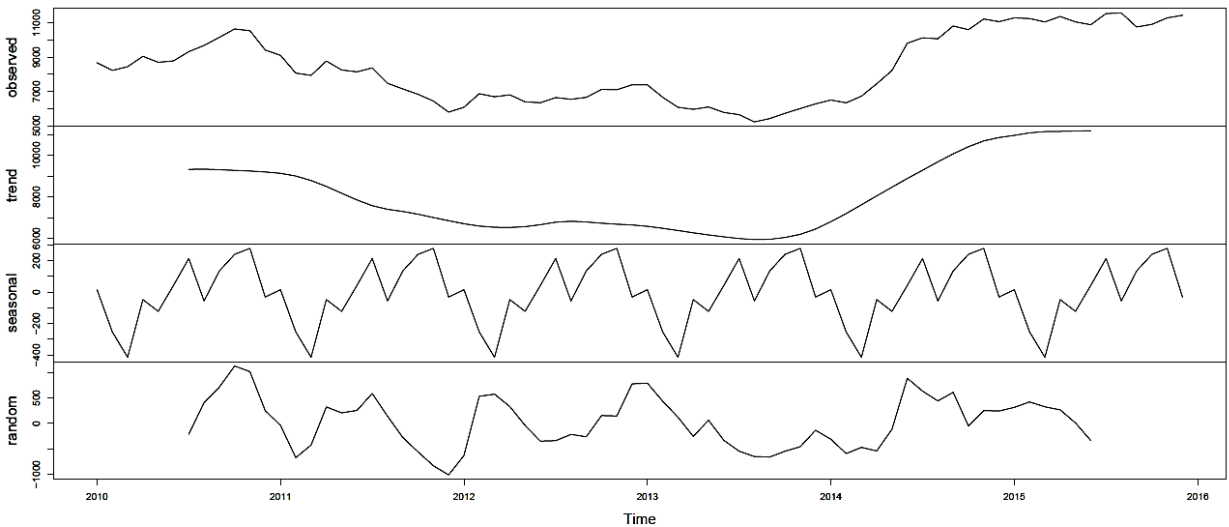

**Figure 4: Decomposition of small cap sector index time series into its trend, seasonal and random components (Jan 2010 – Dec 2015)**

**Table 1: Aggregate value of the Consumer Durables sector index and its components (Jan 2010 – Dec 2015)**

| Year | Month | Aggregate | Trend | Seasonal | Random |
|---|---|---|---|---|---|
| 2010 | Jan | 3890 | | -180 | |
|  | Feb | 4006 | | -185 | |
|  | Mar | 4150 | | -176 | |
|  | Apr | 4512 | | 55 | |
|  | May | 4575 | | 88 | |
|  | Jun | 4518 | | -5 | |
|  | Jul | 5084 | 5248 | 53 | -216 |
|  | Aug | 5658 | 5412 | -78 | 323 |
|  | Sep | 6086 | 5555 | 122 | 410 |
|  | Oct | 6458 | 5705 | 251 | 501 |
|  | Nov | 6742 | 5856 | 182 | 704 |
|  | Dec | 6179 | 6014 | -127 | 291 |
| 2011 | Jan | 6115 | 6175 | -180 | 120 |
|  | Feb | 5735 | 6277 | -185 | -357 |
|  | Mar | 5844 | 6326 | -176 | -306 |
|  | Apr | 6429 | 6351 | 55 | 24 |
|  | May | 6271 | 6330 | 88 | -147 |
|  | Jun | 6620 | 6275 | -5 | 350 |
|  | Jul | 6830 | 6224 | 53 | 553 |
|  | Aug | 6362 | 6227 | -78 | 213 |
|  | Sep | 6563 | 6278 | 122 | 164 |
|  | Oct | 6581 | 6315 | 251 | 14 |
|  | Nov | 6124 | 6335 | 182 | -393 |
|  | Dec | 5463 | 6323 | -127 | -733 |
| 2012 | Jan | 5628 | 6282 | -180 | -474 |
|  | Feb | 6277 | 6261 | -185 | 201 |
|  | Mar | 6522 | 6259 | -176 | 439 |
|  | Apr | 6659 | 6281 | 55 | 323 |
|  | May | 6500 | 6361 | 88 | 518 |
|  | Jun | 6108 | 6518 | -5 | -405 |
|  | Jul | 6353 | 6703 | 53 | -402 |
|  | Aug | 6337 | 6830 | -78 | -416 |
|  | Sep | 6549 | 6895 | 122 | -467 |
|  | Oct | 7126 | 6936 | 251 | -61 |
|  | Nov | 7504 | 7000 | 182 | 322 |
|  | Dec | 7852 | 7071 | -127 | 908 |
| 2013 | Jan | 7663 | 7094 | -180 | 749 |
|  | Feb | 7300 | 7075 | -185 | 410 |
|  | Mar | 7053 | 7026 | -176 | 203 |
|  | Apr | 7115 | 6945 | 55 | 115 |
|  | May | 7574 | 6826 | 88 | 660 |
|  | Jun | 6737 | 6668 | -5 | 74 |
|  | Jul | 6288 | 6497 | 53 | -262 |
|  | Aug | 5936 | 6347 | -78 | -333 |
|  | Sep | 5788 | 6245 | 122 | -579 |
|  | Oct | 5943 | 6190 | 251 | -499 |
|  | Nov | 5822 | 6167 | 182 | -527 |
|      | Dec | 5750  | 6230  | -127 | –353 |
| 2014 | Jan | 5648  | 6395  | -180 | -568 |
|      | Feb | 5718  | 6612  | -185 | -709 |
|      | Mar | 6199  | 6891  | -176 | -515 |
|      | Apr | 6650  | 7205  | 55   | -609 |
|      | May | 7467  | 7523  | 88   | -144 |
|      | Jun | 8357  | 7846  | -5   | 516  |
|      | Jul | 8647  | 8190  | 53   | 404  |
|      | Aug | 8784  | 8572  | -78  | 290  |
|      | Sep | 9616  | 8945  | 122  | 550  |
|      | Oct | 9659  | 9287  | 251  | 121  |
|      | Nov | 9728  | 9575  | 182  | -30  |
|      | Dec | 9615  | 9778  | -127 | -36  |
| 2015 | Jan | 10027 | 9958  | -180 | 249  |
|      | Feb | 10502 | 10156 | -185 | 531  |
|      | Mar | 10373 | 10294 | -176 | 256  |
|      | Apr | 10693 | 10414 | 55   | 224  |
|      | May | 10342 | 10597 | 88   | -343 |
|      | Jun | 10336 | 10798 | -5   | -457 |
|      | Jul | 10985 |       | 53   |      |
|      | Aug | 11208 |       | -78  |      |
|      | Sep | 10498 |       | 122  |      |
|      | Oct | 11671 |       | 251  |      |
|      | Nov | 12097 |       | 182  |      |
|      | Dec | 12075 |       | -127 |      |

### 3.1 Analysis of the Time Series Decomposition Results

Based on the decomposition work on the time series of the two sectors, we make the following important observations:

1. From Table 1, we observe that the seasonal components for the Consumer Durables sector index are positive during the period April-May and September- November, with the highest value occurring in the month of November. The seasonal component is the minimum in the month of February every year. The trend values consistently increased over the period 2010 – 2015 albeit with a sluggish rate. The random component has shown considerable fluctuations in its values. However, the trend is the predominant component in the overall time series.

2. It is natural for the Consumer Durables sector in India to have a dominant seasonal component, as purchase of consumer durable items like air conditioners, refrigerators etc. tend to happen more during the summer period (April-May) and the consumer electronic items like television, micro wave ovens, home theatre systems etc. are sold more during the festive seasons in India which is predominantly during the months of October – November. The companies in the consumer durables sector in India usually run a number of promotion and price discounts schemes during the festive seasons that lead to increased sales of these items, thus leading to a positive seasonal effect during the festive months.

3. From Table 2, the time series for the Small Cap sector also is predominantly guided by its trend component. However, when we look at the strength of the random component values with respect to the overall time series value, we observe that in many months, the presence of the random component is quite strong. This also validates our intuition that the Small Cap sector would have a strong random component in its time series.

Table 2: Aggregate value of the Small Cap sector index and its components
(Jan 2010 – Dec 2015)

| Year | Month | Aggregate | Trend | Seasonal | Random |
|---|---|---|---|---|---|
| 2010 | Jan | 8677 |  | 15 |  |
|  | Feb | 8230 |  | -253 |  |
|  | Mar | 8448 |  | -415 |  |
|  | Apr | 9053 |  | -47 |  |
|  | May | 8702 |  | -123 |  |
|  | Jun | 8785 |  | 43 |  |
|  | Jul | 9324 | 9324 | 214 | -214 |
|  | Aug | 9687 | 9336 | -56 | 408 |
|  | Sep | 10158 | 9308 | 135 | 715 |
|  | Oct | 10647 | 9275 | 241 | 1131 |
|  | Nov | 10544 | 9245 | 279 | 1020 |
|  | Dec | 9420 | 9200 | -31 | 251 |
| 2011 | Jan | 9109 | 9134 | 15 | -40 |
|  | Feb | 8072 | 9003 | -253 | -677 |
|  | Mar | 7942 | 8786 | -415 | -429 |
|  | Apr | 8777 | 8502 | -47 | 323 |
|  | May | 8256 | 8172 | -123 | 207 |
|  | Jun | 8147 | 7850 | 43 | 254 |
|  | Jul | 8377 | 7573 | 214 | 590 |
|  | Aug | 7482 | 7397 | -56 | 141 |
|  | Sep | 7150 | 7295 | 135 | -280 |
|  | Oct | 6838 | 7161 | 241 | -564 |
|  | Nov | 6444 | 7002 | 279 | -836 |
|  | Dec | 5800 | 6849 | -31 | -1018 |
| 2012 | Jan | 6084 | 6702 | 15 | -634 |
|  | Feb | 6871 | 6591 | -253 | 533 |
|  | Mar | 6693 | 6532 | -415 | 576 |
|  | Apr | 6807 | 6523 | -47 | 331 |
|  | May | 6398 | 6563 | -123 | -42 |
|  | Jun | 6348 | 6657 | 43 | -351 |
|  | Jul | 6649 | 6777 | 214 | -342 |
|  | Aug | 6548 | 6823 | -56 | -218 |
|  | Sep | 6659 | 6789 | 135 | -264 |
|  | Oct | 7122 | 6728 | 241 | 153 |
|  | Nov | 7104 | 6681 | 279 | 145 |

|      | Month | | | | |
|------|-------|------|-------|------|------|
|      | Dec   | 7392 | 6644  | -31  | 779  |
| 2013 | Jan   | 7386 | 6579  | 15   | 792  |
|      | Feb   | 6666 | 6482  | -253 | 437  |
|      | Mar   | 6081 | 6376  | -415 | 120  |
|      | Apr   | 5962 | 6266  | -47  | -257 |
|      | May   | 6101 | 6163  | -123 | 61   |
|      | Jun   | 5779 | 6070  | 43   | -334 |
|      | Jul   | 5652 | 5988  | 214  | -550 |
|      | Aug   | 5227 | 5937  | -56  | -654 |
|      | Sep   | 5421 | 5951  | 135  | -665 |
|      | Oct   | 5731 | 6040  | 241  | -550 |
|      | Nov   | 6009 | 6191  | 279  | -461 |
|      | Dec   | 6280 | 6448  | -31  | -137 |
| 2014 | Jan   | 6504 | 6803  | 15   | -314 |
|      | Feb   | 6341 | 7191  | -253 | -597 |
|      | Mar   | 6729 | 7618  | -415 | -474 |
|      | Apr   | 7454 | 8046  | -47  | -544 |
|      | May   | 8228 | 8466  | -123 | -116 |
|      | Jun   | 9815 | 8884  | 43   | 889  |
|      | Jul   | 10132| 9283  | 214  | 635  |
|      | Aug   | 10073| 9687  | -56  | 442  |
|      | Sep   | 10819| 10072 | 135  | 612  |
|      | Oct   | 10604| 10416 | 241  | -52  |
|      | Nov   | 11227| 10697 | 279  | 251  |
|      | Dec   | 11072| 10860 | -31  | 243  |
| 2015 | Jan   | 11294| 10964 | 15   | 315  |
|      | Feb   | 11255| 11085 | -253 | 423  |
|      | Mar   | 11057| 11146 | -415 | 326  |
|      | Apr   | 11375| 11157 | -47  | 266  |
|      | May   | 11059| 11172 | -123 | 9    |
|      | Jun   | 10894| 11191 | 43   | -339 |
|      | Jul   | 11544|       | 214  |      |
|      | Aug   | 11578|       | -56  |      |
|      | Sep   | 10764|       | 135  |      |
|      | Oct   | 10916|       | 241  |      |
|      | Nov   | 11294|       | 279  |      |
|      | Dec   | 11444|       | -31  |      |

In order to investigate further into the behavior of the two time series, we carry out two more experiments. This is driven by our two hypotheses: (i) The Consumer Durables sector displays stronger seasonal characteristics than the Small Cap sector and (ii) The Small Cap sector is dominated by the random component of its time series than the Consumer Durables sector.

Since the absolute values of the time series indices for the two sectors have different scales, it would not make much sense to compare the absolute values of the random and seasonal components of the two time series. Hence, we prepared four text files containing the percentage values of the random and the seasonal components with respect to the overall time series values for both the sectors over the period January 2010 – December 2015. From these four text files, we created four time series variables in R using the *ts( )* function in the *TTR* package. Using the two seasonal components of the time series (one each for the two sectors), we have created a multiple line plot so that the seasonal components for the two sectors can be visually compared. The same exercise is repeated for the random component time series.

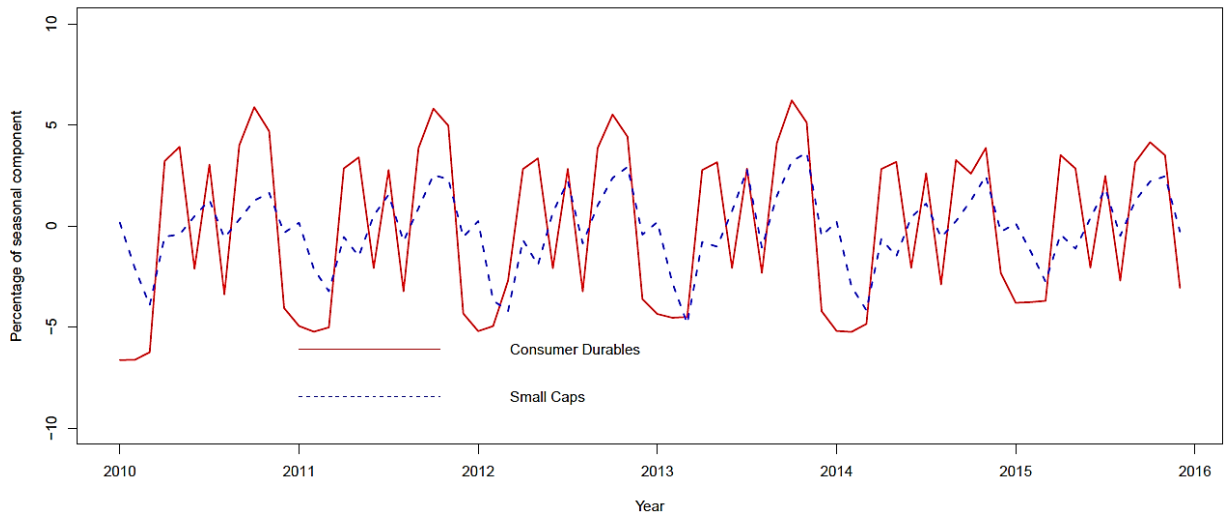

**Figure 5: Comparison of the seasonal components of the consumer durable and the small caps sector (Jan 2010 – Dec 2015)**

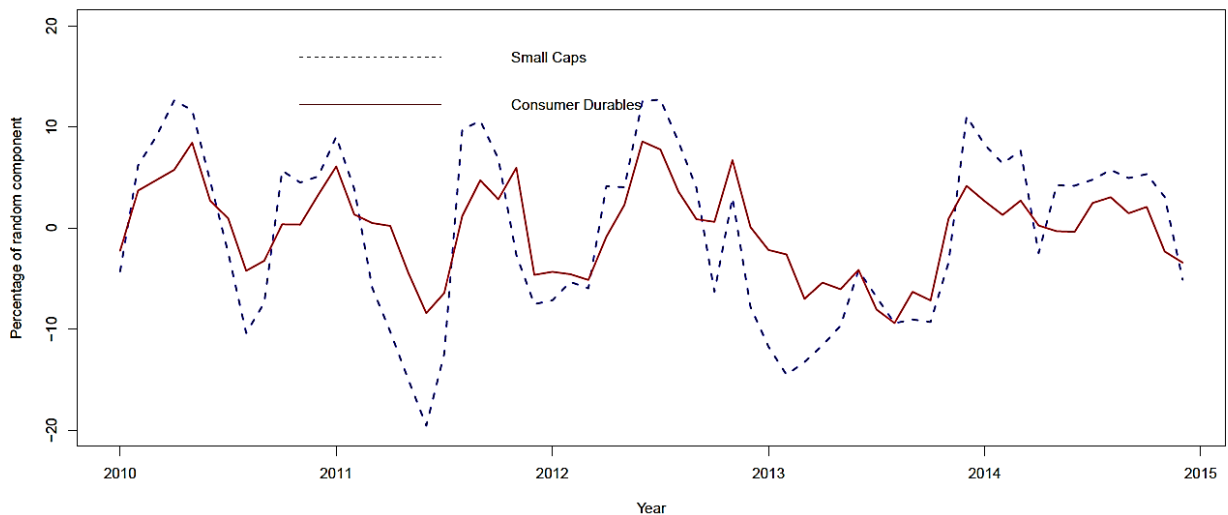

**Figure 6: Comparison of the random components of the consumer durable and the small caps sector (Jan 2010 – Dec 2015)**

Figure 5 depicts the comparison of the percentage of the seasonal components in the overall time series for the Consumer Durables and the Small Cap sector. It can be easily observed that the crests and the troughs of the Consumer Durables sector are much bigger than those of the Small

Cap sector. Hence, the results validate our hypothesis (i) – Consumer Durable sector exhibits more seasonality than the Small Cap sector.

Figure 6 presents the comparison of the percentage of the random components in the overall time series for the Consumer Durables and the Small Cap sector. It not difficult to observe that curve for the Small Cap sector has longer amplitude in fluctuations compared to its Consumer Durable counterpart. Hence, the results validate our hypothesis (ii) – Small Cap sector exhibits stronger random component in its time series than the Consumer Durable sector.

4. Proposed Frameworks of Time Series Forecasting and Analysis

In this Section, we discuss some methods that we have applied on the Consumer Durables time series data and the Small Cap time series data for making robust forecasting and for a better understanding of the relative contributions of the constituents (i.e., the trend, seasonal and random components) of a time series. We present five different approaches in forecasting and one method for determining the relative strengths of the trend and seasonality components in a time series.

**Method 1:** The time series data of the Consumer Durables and the Small Cap sectors for the period January 2010 to December 2014 is used for building the forecasting model. The *HoltWinters( )* function in R with changing slope in trend and presence of seasonal components is used to forecast the monthly indices for both the sectors for each month of 2015. The forecasted values are compared with the actual values of the indices and the error of forecast is computed for every month of 2015 for both the sectors. Note that in the approach, the forecast is made in December 2014 for every month of 2015. Therefore, the forecast horizon in the approach is 12 months.

**Method II:** In contrast to Method I, in this approach, the forecasting for each month of 2015 for both the sectors are done on the basis of time series data from January 2010 till the end of the previous month for which the forecast is made. For example, in order to forecast for the month of April 2015, time series data for the period January 2010 to March 2015 are used to build the forecasting model. Hence, this method uses a forecast horizon 1 month. The *HoltWinters( )* function in R with changing slope in trend and presence of seasonal component is used in forecasting. The errors in forecast are computed for each sector for every month of 2015 in the same manner as in Method I.

**Method III:** The fundamental objective of this approach of forecasting is to investigate how effectively we can forecast the aggregate of the trend and the seasonal components of a times series. Since, the random components in a time series are impossible to predict, we devise an approach of forecasting using the trend and seasonal components of a time series. In this method, we first use the time series data for both the Consumer Durables and the Small Cap sectors for the period January 2010 to December 2014 and decompose both the time series into their respective trend and seasonal and random components. As we have seen in Section 2, the decomposition yields the trend component from July 2010 to June 2014 for each time series, since values for the first six months and last six months are truncated for computations of 12 months' moving averages. Using the computed trends values for both the sectors for the period July 2010 to June 2014, we forecast the trends values for both the sectors for the period January 2015 to June 2015, using *HoltWinters( )* function in R with changing slope in the trend component and a seasonal component (Coghlan, 2015). These forecasted trend values are added to the corresponding monthly seasonal components which were obtained from the decomposition of the time series data for the period January 2010 to December 2014 for both the sectors. These aggregate values of the trend and the seasonal components now constitute our forecasted aggregate trend and seasonal values for both the sectors for the period January 2015 to June 2015. In order to compute the actual aggregates of the trend and seasonal components, we use the time series for both the sectors for the period January 2010 to December 2015 and decompose both the time series into their trend, seasonal and random components. After decomposition, we compute the aggregate of the actual trend and the actual seasonal values for both the time series for the period January 2015 to June 2015. The errors of forecasts for each month for both the sectors are also computed.

**Method IV:** In this approach, we have used *Auto Regressive Integrated Moving Average* (ARIMA) technique (Coghlan, 2015) for forecasting. Two ARIMA forecast models (one model each for the two sectors – Consumer Durables and Small Cap) are built using the two time series for the Consumer Durables and the Small Cap sectors for the period January 2010 – December 2014. Based on each of these two time series data, we first derive the three parameters of the *Auto Regressive Moving Average* (ARMA) mode, i.e. the *Auto Regression* parameter (p), the *Difference* parameter (d), and the *Moving Average* parameter (q) for both the time series. The values of the three parameters are used to develop the ARIMA models for the two sectors. Finally, the two ARIMA models are used to predict the time series values of the respective sectors for each month of 2015. Since the forecasting for all the months of 2015 is being made in December 2014, the forecast horizon in this ARIMA approach is 12 months.

**Method V:** In this approach, forecasting is done for both the sectors using two ARIMA models (one each for the two sectors) as in Method IV. However, in contrast to Method IV, where we used a forecast horizon of 12 months, in this approach, we have used a forecast horizon of 1 month. Therefore, for forecasting the time series value for each month in 2015, the training data set for building the ARIMA model included time series data from January 2010 till the last month for which the forecast was made. For example, if we need to forecast the time series value for the Consumer Durables sector for the month of June 2015, the training data set for building the ARIMA model would include time series data from January 2010 till May 2015. It is important to note here that since the training data set for the ARIMA model in this approach is constantly changing (due to inclusion of newer data), it is mandatory to evaluate the ARIMA parameters every time before the forecasting is made for each month of 2015.

**Method VI:** Ideally, in a time series, both the trend and the seasonal components would vary over time. The variation of the random component is also there, However, the variations of the random component is difficult to model and hence the focus of our forecasting approaches is on the variations of the trend and seasonal components. In this approach, we investigate how the seasonal components in the time series vary with time for both the sectors under considerations, e.g., the Consumer Durables sector and the Small Cap sector. For this purpose, we first consider the time series of both the sectors for the period January 2010 to December 2014. Each of these two time series is decomposed into its trend, seasonal and random components and the aggregate values of the trend and the seasonal components for the period July 2010 to June 2014 are computed. It may be noted that the aggregates could not be computed for the periods January 2010 to June 2010 and July 2014 to December 2014 due to truncation of the trend values. Next, we remove the time series values for the period January 2010 to December 2010 from the data under investigation and insert the time series values for the period January 2015 to December 2015. In other words, we now concentrate on the time series values for the period January 2011 to December 2015 for both the sectors. As in the previous case, we again compute the aggregate of the trend and the seasonal components for period July 2010 to June 2014 for both the sectors based on the new time series from January 2011 to December 2015. Since the seasonal component values are expected to change from 2010 to 2015, in order to have an idea about the change in the aggregate values of the trend and the seasonal components, we compute the percentage of deviation of the computed aggregate of the trend and the seasonal components for both the sector for each month during the period of June 2011 to July 2014, computed based on the two time series (January 2010 – December 2014 and January 2011 – December 2015). In the event of appreciable changes in the seasonal component values, we expect large values of the percentage deviations.

## 5. Forecasting Results and Analysis

As discussed in Section 4, we have applied five forecasting methods and time series analysis technique on both the Consumer Durables and the Small Cap sectors' time series. In this Section, we present the detailed results and a critical analysis of the relative merits and demerits of each of the forecasting and analysis framework.

**Method I:** As discussed in Section 5, for both the sectors, we make forecast for each month of 2015 based on time series data from January 2010 to December 2014. *HoltWinters( )* function in R library *forecast* has been used with changing slope in trend (i.e., varying trend) and a seasonal component. The *forecast horizon* in the *HoltWinters* model has been chosen to be 12 months for both the sectors so that the forecasted values for all months of 2015 can be obtained. The results of forecasting using Method I for the Consumer Durables sector and the Small Cap sector are presented in Table 3 and Table 4 respectively.

**Table 3: Results of Method I of forecasting for Consumer Durables sector
(Jan 2015 – Dec 2015)**

| Month | Actual Value | Forecasted Value | Error Percentage |
|---|---|---|---|
| (A) | (B) | (C) | (C-B)/B *100 |
| Jan | 10027 | 9451 | 5.74 |
| Feb | 10502 | 9146 | 12.91 |
| Mar | 10373 | 9166 | 11.63 |
| Apr | 10693 | 9647 | 9.78 |
| May | 10342 | 9839 | 4.86 |
| Jun | 10336 | 10076 | 2.52 |
| Jul | 10985 | 9974 | 9.20 |
| Aug | 11208 | 10144 | 9.49 |
| Sep | 10498 | 10649 | 1.44 |
| Oct | 11671 | 10990 | 5.83 |
| Nov | 12097 | 11197 | 7.44 |
| Dec | 12075 | 10859 | 10.07 |

**Table 4: Results of Method I of forecasting for Small Cap sector
(Jan 2015 – Dec 2015)**

| Month | Actual Value | Forecasted Value | Error Percentage |
|---|---|---|---|
| (A) | (B) | (C) | (C-B)/B *100 |
| Jan | 11294 | 10790 | 4.46 |
| Feb | 11255 | 10208 | 9.31 |
| Mar | 11057 | 10105 | 8.61 |
| Apr | 11375 | 10962 | 3.63 |
| May | 11059 | 11396 | 3.05 |
| Jun | 10894 | 11885 | 9.10 |
| Jul | 11544 | 11942 | 3.45 |
| Aug | 11578 | 12000 | 3.64 |
| Sep | 10764 | 12729 | 18.26 |
| Oct | 10916 | 13354 | 22.33 |
| Nov | 11294 | 13904 | 23.11 |
| Dec | 11444 | 13564 | 18.52 |

**Observations:** We observe from Table 3 that the forecasted values closely match the actual values of the Consumer Durables sector even when the forecast horizon is long (12 months). This clearly shows that *HoltWinters* model with changing trend and additive seasonal components is effective in forecasting the Consumer Durable sector time series values for the period 2010 -2015. The error values exceeded 10% threshold for the months of February, March and December 2015. For the month of February, there was an unexpected increase in the time series value compared to its previous value in January 2015. In March 2015, on the other hand, there was a decrease in the time series index which was also not expected. In December 2015, there was a fall in time series index which was not in tune with the increasing trends in its previous values. The error values in Table 4 indicate that the forecasted values for the Small

Caps sector also very closely match its actual values except for the period September – December 2015. The Small Caps time series behaved in a very unpredictable way during this period. A careful look at Table 2 will make it clear that the time series has undergone alternate rise and fall during this period and it was impossible for *HoltWinters* model with a forecast horizon of 12 months to effectively capture this behavior of the time series which resulted in higher values in forecasting errors.

**Method II:** For both the sectors, we have used *HoltWinters( )* function in R with additive seasonal component and a trend with a changing slope. However, in contrast with Method I, the forecast horizon in this method has been chosen to be 1 month. In other words, forecasting is made for each month of 2015 for both the sectors by taking into account time series data from January 2010 till the previous month for which forecasting is being made. Since this approach uses a very small horizon of forecast, it is likely that this method will be able to capture any possible change in trend and seasonal components more effectively than Method I. However, if there is a continuous rise and fall in the time series values, this method may yield worse results compared to those obtained in Method I. The results of forecasting using Method II for the Consumer Durables sector and the Small Cap sector are presented in Table 5 and Table 6 respectively.

**Table 5: Results of Method II of forecasting for Consumer Durables sector (Jan 2015 – Dec 2015)**

| Month (A) | Actual Value (B) | Forecasted Value (C) | Error Percentage (C-B)/B *100 |
|---|---|---|---|
| Jan | 10027 | 9451 | 5.74 |
| Feb | 10502 | 9655 | 8.07 |
| Mar | 10373 | 10419 | 0.44 |
| Apr | 10693 | 10851 | 1.48 |
| May | 10342 | 10916 | 5.55 |
| Jun | 10336 | 10645 | 2.99 |
| Jul | 10985 | 10274 | 6.47 |
| Aug | 11208 | 11035 | 1.54 |
| Sep | 10498 | 11695 | 11.40 |
| Oct | 11671 | 11001 | 5.74 |
| Nov | 12097 | 11783 | 2.60 |
| Dec | 12075 | 11767 | 2.55 |

**Observations:** We observe from Table 5 that the forecasted values very closely match with the actual values for the Consumer Durable sector. Except for the month of September 2015, the forecast error values have never exceeded the threshold of 10%. The higher error value for the month of September may be attributed to the sudden and unexpected fall in the time series value for that month. The time series was consistently on the rise over the previous few months, and since the forecast horizon is 1 month, *HoltWinters* method expected an increase in the time series value following the increasing trend of the time series. However, the time series value actually decreased and that resulted into a higher value in foresting error. The error values in Table 6 indicate that the forecasted values for the Small Cap sector also very closely match its actual values except for the month of September 2015. The sudden rise in the error value in September 2015 for the Small Caps sector can again be attributed to the sudden decrease in the time series value during that month which was inconsistent with the increasing trend of the time series in the previous few months.

Table 6: Results of Method II of forecasting for Small Cap sector (Jan 2015 – Dec 2015)

| Month | Actual Value | Forecasted Value | Error Percentage |
|---|---|---|---|
| (A) | (B) | (C) | (C-B)/B *100 |
| Jan | 11294 | 10790 | 4.46 |
| Feb | 11255 | 10669 | 5.21 |
| Mar | 11057 | 11160 | 0.93 |
| Apr | 11375 | 12021 | 5.68 |
| May | 11059 | 11982 | 8.35 |
| Jun | 10894 | 11695 | 7.35 |
| Jul | 11544 | 10960 | 5.06 |
| Aug | 11578 | 11445 | 1.16 |
| Sep | 10764 | 12233 | 13.65 |
| Oct | 10916 | 11504 | 5.39 |
| Nov | 11294 | 11504 | 1.86 |
| Dec | 11444 | 10844 | 5.24 |

**Method III:** In Section 5, we have already discussed the approach followed in this method. We have used the time series data of the Consumer Durables sector from January 2010 to December 2015 to compute the actual values of the trend and the seasonal components. However, since the actual values of trend component are not available for the period July 2015 – December 2015, we concentrate only on the period January 2015 to June 2015 for the purpose of forecasting. The actual trend and seasonal component values and their aggregated monthly values are noted in Columns *B*, *C* and *D* respectively in Table 7. Now, using the time series data for the period January 2010 to December 2014, the trend and the seasonal components are recomputed. Since the trend values during July 2014 to December 2014 will not be available after this computation, we make a forecast for the trend values for the period January 2015 to June 2015 using *HoltWinters* forecasting model with a changing trend and an additive seasonal component. The forecasted trend values and the past seasonal component values and their corresponding aggregate values are noted in columns *E*, *F* and *G* respectively in Table 7. The error values are also computed and recorded in the rightmost column of the Table 7. For the Small Cap sector time series, the same method has been followed and the results are recorded in Table 8.

**Observation:** The results obtained using Method III for the Consumer Durables and the Small Caps sectors are presented in Table 7 and Table 8 respectively. From Table 7, we observe that the error values have consistently increased from 2.29% in October 2014 to 19.42% in June 2015. Considering the fact that the trend is forecasted over a long period of 12 months (forecasting for July 2014 – June 2015 being done at the end of June 2014) and since the trend component of the Consumer Durables sector had been sluggish over the period of forecast, the forecasting accuracies obtained in Method III can be considered quite satisfactory for the Consumer Durables sector. The fact that the actual trend was not able to keep its pace intact with its forecasted values is evident from the values in the column *B* and the corresponding values in the column *E* of the Table 7.

Since the trend component of the Small Cap sector grew even more sluggishly as compared to the Consumer Durables sector over the period July 2014 – June 2015, the forecast errors for Small Caps sector using Method III have been higher. It is evident from Table 8 that forecast errors has consistent increased from 0.88% in July 2014 to 35.05% in June 2015. The sluggish rate of increase in the actual trend component as compared to the forecasted trend component using the *HoltWinters( )* function with changing slope in trend and an additive seasonal component is evident from the values in the column *B* and the corresponding values in the column *E* of the Table 8. The extreme sluggish rate of growth in the trend component in the Small Cap sector during the period of July 2014 to June 2015 has rendered Method III not a very effective method of forecasting for the Small Cap sector.

**Table 7: Results of Method III of forecasting for Consumer Durables sector (July 2014 – June 2015)**

| Month | Actual Trend | Actual Seasonal | Actual (Trend + Seasonal) | Forecasted Trend | Past Seasonal | Forecasted (Trend + Seasonal) | % Error |
|---|---|---|---|---|---|---|---|
| A | B | C | D | E | F | G | (G-D)/D *100 |
| Jul 2014 | 8190 | 53 | 8243 | 8192 | -12 | 8180 | 0.76 |
| Aug 2014 | 8572 | -78 | 8494 | 8565 | -113 | 8452 | 0.49 |
| Sep 2014 | 8945 | 122 | 9067 | 9030 | 21 | 9051 | 0.18 |
| Oct 2014 | 9287 | 251 | 9538 | 9498 | 258 | 9756 | 2.29 |
| Nov 2014 | 9575 | 182 | 9757 | 9936 | 226 | 10162 | 4.15 |
| Dec 2014 | 9778 | -127 | 9651 | 10327 | -81 | 10246 | 6.17 |
| Jan 2015 | 9958 | -180 | 9778 | 10842 | -206 | 10636 | 8.77 |
| Feb 2015 | 10156 | -185 | 9971 | 11315 | -281 | 11034 | 10.66 |
| Mar 2015 | 10294 | -176 | 10118 | 11713 | -204 | 11509 | 13.75 |
| Apr 2015 | 10414 | 55 | 10469 | 12081 | 35 | 12116 | 15.73 |
| May 2015 | 10597 | 88 | 10685 | 12423 | 210 | 12633 | 18.23 |
| Jun 2015 | 10798 | -5 | 10793 | 12743 | 146 | 12889 | 19.42 |

**Table 8: Results of Method III of forecasting for Small Cap sector (July 2014 – June 2015)**

| Month | Actual Trend | Actual Seasonal | Actual (Trend + Seasonal) | Forecasted Trend | Past Seasonal | Forecasted (Trend + Seasonal) | % Error |
|---|---|---|---|---|---|---|---|
| A | B | C | D | E | F | G | (G-D)/D *100 |
| Jul 2014 | 9283 | 214 | 9497 | 9293 | 120 | 9413 | 0.88 |
| Aug 2014 | 9687 | -56 | 9631 | 9845 | -102 | 9743 | 1.16 |
| Sep 2014 | 10072 | 135 | 10207 | 10443 | 47 | 10490 | 2.77 |
| Oct 2014 | 10416 | 241 | 10657 | 10996 | 319 | 11315 | 6.17 |
| Nov 2014 | 10697 | 279 | 10976 | 11497 | 281 | 11778 | 7.31 |
| Dec 2014 | 10860 | -31 | 10829 | 11970 | -27 | 11943 | 10.29 |
| Jan 2015 | 10964 | 15 | 10979 | 12840 | 2 | 12842 | 16.97 |
| Feb 2015 | 11085 | -253 | 10832 | 13394 | -294 | 13100 | 20.94 |
| Mar 2015 | 11146 | -415 | 10731 | 13852 | -431 | 13421 | 25.07 |
| Apr 2015 | 11157 | -47 | 11110 | 14247 | -49 | 14198 | 27.79 |
| May 2015 | 11172 | -123 | 11049 | 14605 | -60 | 14545 | 31.64 |
| Jun 2015 | 11191 | 43 | 11234 | 14979 | 193 | 15172 | 35.05 |

**Method IV:** As pointed out in Section 4, we have applied *Auto Regressive Integrated Moving Average* (ARIMA) technique for the purpose of forecasting on the Consumer Durables and the Small Cap time series data. We have exploited the power of the *auto.arima( )* function defined in the *forecast* package in R for determining the values of the ARIMA parameters for the time series of the two sectors (Coghlan, 2015). For finding the values of the ARIMA parameters for both the time series, we have used the time series for the Consumer Durables sector and the time series for the Small Cap sector for the period January 2010 to December 2014. For the Consumer Durable sector, applying the *auto.arima ( )* function on the time series, we obtain the ARIMA parameters: *Auto Regression* parameter (p) = 0, *Difference* parameter (d) = 1, *Moving Average* parameter (q) = 1. Therefore, the Consumer Durables sector time series for the period January 2010 – December 2014 is designed as an *Auto Regressive Moving Average* (ARMA) model - ARMA(0, 1, 1). From the ARMA(0, 1, 1) model, the corresponding ARIMA model is constructed using the *arima( )* function in R with the two parameters as: (i) Consumer Durables sector time series, (ii) the order of the ARMA for the time series, i.e., (0, 1, 1). Using the resultant ARIMA model, we call the function *forecast.Arima( )* with parameters: (i) the ARIMA model and (ii) the time horizon of forecast. In this method (i.e., Method IV), we make the forecast for each month of the year 2015 based on the time series for the period January 2010 to December 2014, resulting in a forecast horizon of 12 months. The errors in forecasting are also computed. The same approach is also followed for the Small Cap sector. However, the ARMA parameters for the Small Caps time series were found to be (1, 1, 0). The ARIMA model for the Small Caps sector was built according to these values. The results of forecasting using Method IV for the Consumer Durables sector and the Small Cap sector are presented in Table 9 and Table 10 respectively.

**Observation:** From Table 9, it may be observed that the forecast error values are very low for the Consumer Durable sector with the ARIMA method even using a large forecast horizon of 12 months. The three months (May, June and September 2015) for which the error values exceeded the threshold of 10% mark, the time series exhibited unexpected fall as compared to its previous values and hence it was difficult for the ARIMA model with a long forecast horizon of 12 months to predict this behavior. The error values in Table 10 indicate that the Small Cap sector time series for 2015 had a very close fit with the ARIMA model even with a long forecast horizon of 12 months. The maximum error in forecast being less than 5%, the model has provided an excellent framework of forecast for the Small Cap sector. The Small Cap time series for the year 2015 has remained consistent with no abrupt and sudden increase/decrease in its values. This has allowed ARIMA, even with a long forecast horizon of 12 months, to provide a very accurate framework for forecasting.

**Method V:** In this method, we have utilized ARIMA model for forecasting with a forecast horizon of 1 month for both the sectors. The methodology used for constructing the ARIMA models has been the same as it was in Method IV. However, since the model deploys a training data set that is constantly increasing in size due to inclusion of new data, it is mandatory to re-evaluate the ARIMA parameters every time before the *forecast.Arima( )* function is used for the purpose of forecasting. In other words, before forecasting is made for each month of 2015, we compute the values of the ARIMA parameters, and build a new ARIMA model for forecasting. The results of forecasting using Method V for the Consumer Durables sector and the Small Cap sector are presented in Table 11 and Table 12 respectively.

**Observations:** From Table 11, it may be observed that the forecast errors are very low for the Consumer Durables sector with the ARIMA model using a low forecast horizon of 1 month. This is expected as the small forecast horizon allows the ARIMA model to capture the behavior of the time series more effectively. However, if there is a sharp change in the time series values (i.e. abrupt increase/decrease in the time series values), the ARIMA model with small forecast horizon of 1 month may perform poorly since it assigns highest weight to the last observation. The higher value of forecast error of 12.47% for the month of October 2015 for the Consumer Durables sector may be attributed to this reason. The Consumer Durables sector time series had a fall in its value from August to September which was against the increasing trend of the time series over the last few months. This resulted in a moderate value of 6.77% of the forecast error for the month of September 2015. If this fall continued in the month of October 2015, ARIMA would have provided a very low value of the forecast error. However, in the month of October 2015, the time series exhibited an increase in its value resulting in a high value of the error rate. Since the time series of the Small Cap sector did not exhibit any abrupt increase or fall in its values in the year 2015, ARIMA model with a forecast horizon of 1 month has produced consistently low error rates in forecasting as evident from Table 12.

**Table 9: Results of Method IV of forecasting for Consumer Durables sector (Jan 2015 – Dec 2015)**

| Month | Actual Value | Forecasted Value | Error Percentage |
|---|---|---|---|
| (A) | (B) | (C) | (C-B)/B *100 |
| Jan | 10027 | 10232 | 2.04 |
| Feb | 10502 | 10743 | 2.29 |
| Mar | 10373 | 11095 | 6.96 |
| Apr | 10693 | 11381 | 6.43 |
| May | 10342 | 11628 | 12.43 |
| Jun | 10336 | 11849 | 14.63 |
| Jul | 10985 | 12050 | 9.70 |
| Aug | 11208 | 12237 | 9.18 |
| Sep | 10498 | 12411 | 18.22 |
| Oct | 11671 | 12576 | 7.75 |
| Nov | 12097 | 12732 | 5.24 |
| Dec | 12075 | 12881 | 6.67 |

**Table 10: Results of Method IV of forecasting for Small Cap sector (Jan 2015 – Dec 2015)**

| Month | Actual Value | Forecasted Value | Error Percentage |
|---|---|---|---|
| (A) | (B) | (C) | (C-B)/B *100 |
| Jan | 11294 | 11030 | 2.34 |
| Feb | 11255 | 11018 | 2.11 |
| Mar | 11057 | 11015 | 0.38 |
| Apr | 11375 | 11014 | 3.17 |
| May | 11059 | 11014 | 0.41 |
| Jun | 10894 | 11014 | 1.10 |
| Jul | 11544 | 11014 | 4.59 |
| Aug | 11578 | 11014 | 4.87 |
| Sep | 10764 | 11014 | 2.32 |
| Oct | 10916 | 11014 | 0.90 |
| Nov | 11294 | 11014 | 2.48 |
| Dec | 11444 | 11014 | 3.76 |

**Summary of Performance:** In Table 13 and Table 14, we have summarized the performance of the five forecasting approaches that we have discussed so far for the Consumer Durables sector and the Small Cap sector respectively. For the purpose of comparison, we have chosen four metrics: (i) minimum (min) error rate, (ii) maximum (max) error rate, (iii) mean error rate, and (iv) standard deviation (sd) of error rates.

As observed from Table 13, for the Consumer Durables sector, Method V that used ARIMA with a forecast horizon of 1 month, has performed best in two metrics: (i) min error rate and (ii) mean error rate. Method II that used *HoltWinters* forecasting model with a forecast horizon of 1 month has performed best in the other two metrics: (i) max error rate and (ii) sd of error rates. Method I that used *HoltWinters* forecasting models with a forecast horizon of 12 months has performed next with a mean error rate of 7.58 and an sd of error rate of 3.56. Method III that used aggregate of trend and seasonal components has been ranked at fourth position based on the mean error value of 8.38 and Method IV that used ARIMA with a forecast horizon of 12 months has performed worst with a mean error value of 8.46. It is evident, that for the Consumer Durables time series in the year 2015, forecasting methods that used small forecast horizons produced better results. This is due to the fact that the time series did not exhibit abrupt increase/decrease in its values over successive months over the period of forecast.

From Table 14, for the Small Cap sector, it is evident that the Method IV that used ARIMA with a forecast horizon of 12 months have performed best in all the four metrics: (i) min, (ii) max, (iii) mean, and (iv) sd of error rates. Method V that used ARIMA with a forecast horizon of 1 month has been the next in terms of performance with respect to all the four metrics. Since the time series had exhibited changes in its behavior in quite a number of instances in 2015, ARIMA with forecast horizon of 1 month could not provide the best results. However, since the magnitude of those changes were very nominal, the ARIMA with a long horizon of 12 months produced best forecasting results, and the ARIMA with a forecast horizon of 1 month providing good results too. Method II using *HoltWinters( )* function with a forecast horizon of 1 month and Method I using *HoltWinters( )* function with a forecast horizon of 12 months were next in terms of their mean percentage error values. As expected, Method III that used aggregate of the forecasted trend and seasonal values, produced the worst forecasting results for the Small Cap sector. The trend component of the Small Cap sector has been very slow in its growth resulting in a large gap between actual trend values and the forecasted trend values for the year 2015. This gap is responsible for the high mean percentage error (15.50) in forecasting in Method III for the Small Cap sector.

**Table 11: Results of Method V of forecasting for Consumer Durables sector (Jan 2015 – Dec 2015)**

| Month (A) | Actual Value (B) | Forecasted Value (C) | Error Percentage (C-B)/B *100 |
|---|---|---|---|
| Jan | 10027 | 10231 | 2.03 |
| Feb | 10502 | 10218 | 2.70 |
| Mar | 10373 | 10615 | 2.33 |
| Apr | 10693 | 10272 | 3.94 |
| May | 10342 | 10861 | 5.02 |
| Jun | 10336 | 10214 | 1.18 |
| Jul | 10985 | 10334 | 5.93 |
| Aug | 11208 | 11206 | 0.02 |
| Sep | 10498 | 11209 | 6.77 |
| Oct | 11671 | 10216 | 12.47 |
| Nov | 12097 | 12043 | 0.45 |
| Dec | 12075 | 12108 | 0.27 |

**Table 12: Results using Method V of forecasting for Small Cap (Jan 2015 – Dec 2015)**

| Month (A) | Actual Value (B) | Forecasted Value (C) | Error Percentage (C-B)/B *100 |
|---|---|---|---|
| Jan | 11294 | 11030 | 2.34 |
| Feb | 11255 | 11844 | 5.23 |
| Mar | 11057 | 11245 | 1.70 |
| Apr | 11375 | 11004 | 3.26 |
| May | 11059 | 11459 | 3.62 |
| Jun | 10894 | 10978 | 0.77 |
| Jul | 11544 | 10852 | 5.99 |
| Aug | 11578 | 11706 | 1.11 |
| Sep | 10764 | 11586 | 7.64 |
| Oct | 10916 | 10668 | 2.27 |
| Nov | 11294 | 11027 | 2.36 |
| Dec | 11444 | 11296 | 1.29 |

**Table 13: Performance of the forecasting methods for the Consumer Durables sector**

| Metrics / Methods | Min Error | Max Error | Mean Error | SD of Errors |
|---|---|---|---|---|
| Method 1 | 1.44 | 12.91 | 7.58 | 3.56 |
| Method II | 0.44 | 11.40 | 4.55 | 3.20 |
| Method III | 0.18 | 19.42 | 8.38 | 7.10 |
| Method IV | 2.04 | 18.22 | 8.46 | 4.78 |
| Method V | 0.02 | 12.47 | 3.59 | 3.58 |

**Table 14: Performance of the forecasting methods for the Small Cap sector**

| Metrics / Methods | Min Error | Max Error | Mean Error | SD of Errors |
|---|---|---|---|---|
| Method 1 | 3.05 | 23.11 | 10.62 | 7.78 |
| Method II | 0.93 | 13.65 | 5.36 | 3.47 |
| Method III | 0.88 | 35.05 | 15.50 | 12.36 |
| Method IV | 0.38 | 4.87 | 2.37 | 1.52 |
| Method V | 0.77 | 7.64 | 3.13 | 2.14 |

**Table 15: Computation Results using Method VI**
**(Structural analysis of trend and seasonal components of the Consumer Durables sector time series for the Period: July 2011 – June 2014)**

| Year | Month | Computation 1 (Based on 2010-2014) | | | Computation 2 (Based on 2011-2015) | | | % Variation |
|---|---|---|---|---|---|---|---|---|
| | | Trend | Seasonal | Sum | Trend | Seasonal | Sum | (F - C)/C *100 |
| | | A | B | C = A + B | D | E | F = D + E | |
| 2011 | Jul | 6224 | -12 | 6212 | 6224 | 142 | 6366 | 2.48 |
| | Aug | 6227 | -113 | 6114 | 6227 | -123 | 6104 | 0.16 |
| | Sep | 6278 | 21 | 6299 | 6278 | 54 | 6332 | 0.52 |
| | Oct | 6315 | 258 | 6573 | 6315 | 161 | 6476 | 1.48 |
| | Nov | 6335 | 226 | 6561 | 6335 | 42 | 6377 | 2.80 |
| | Dec | 6323 | -81 | 6242 | 6323 | -164 | 6159 | 1.33 |
| 2012 | Jan | 6282 | -206 | 6076 | 6282 | 175 | 6107 | 0.51 |
| | Feb | 6261 | -281 | 5980 | 6261 | -61 | 6200 | 3.67 |
| | Mar | 6259 | -204 | 6055 | 6259 | -65 | 6194 | 2.30 |
| | Apr | 6281 | 35 | 6316 | 6281 | 84 | 6365 | 0.78 |
| | May | 6361 | 210 | 6571 | 6361 | 160 | 6521 | 0.76 |
| | Jun | 6518 | 146 | 6664 | 6518 | -57 | 6461 | 3.05 |
| | Jul | 6703 | -12 | 6691 | 6703 | 142 | 6845 | 2.31 |
| | Aug | 6830 | -113 | 6717 | 6830 | -123 | 6707 | 0.15 |
| | Sep | 6895 | 21 | 6916 | 6895 | 54 | 6949 | 0.48 |
| | Oct | 6936 | 258 | 7194 | 6936 | 161 | 7097 | 1.35 |
| | Nov | 7000 | 226 | 7226 | 7000 | 42 | 7042 | 2.55 |
| | Dec | 7071 | -81 | 6990 | 7071 | -164 | 6907 | 1.19 |
| 2013 | Jan | 7094 | -206 | 6888 | 7094 | 175 | 6919 | 0.45 |
| | Feb | 7075 | -281 | 6794 | 7075 | -61 | 7014 | 3.24 |
| | Mar | 7026 | -204 | 6822 | 7026 | -65 | 6961 | 2.04 |
| | Apr | 6945 | 35 | 6980 | 6945 | 84 | 7029 | 0.70 |
| | May | 6826 | 210 | 7036 | 6826 | 160 | 6986 | 0.71 |
| | Jun | 6668 | 146 | 6814 | 6668 | -57 | 6611 | 2.98 |
| | Jul | 6497 | -12 | 6485 | 6497 | 142 | 6639 | 2.38 |
| | Aug | 6347 | -113 | 6234 | 6347 | -123 | 6224 | 0.16 |
| | Sep | 6245 | 21 | 6266 | 6245 | 54 | 6299 | 0.53 |
| | Oct | 6190 | 258 | 6448 | 6190 | 161 | 6351 | 1.50 |
| | Nov | 6167 | 226 | 6393 | 6167 | 42 | 6209 | 2.89 |
| | Dec | 6230 | -81 | 6149 | 6230 | -164 | 6066 | 1.35 |
| 2014 | Jan | 6395 | -206 | 6189 | 6395 | -175 | 6220 | 0.50 |
| | Feb | 6612 | -281 | 6331 | 6612 | -61 | 6551 | 3.47 |
| | Mar | 6891 | -204 | 6687 | 6891 | -65 | 6826 | 2.08 |
| | Apr | 7205 | 35 | 7240 | 7205 | 84 | 7289 | 0.68 |
| | May | 7523 | 210 | 7733 | 7523 | 160 | 7683 | 0.65 |
| | Jun | 7846 | 146 | 7992 | 7846 | -57 | 7789 | 2.54 |

**Table 16: Computation Results using Method VI**
**(Structural analysis of trend and seasonal components of the Small Cap sector time series for the Period: July 2011 – June 2014)**

| Year | Month | Computation 1 (Based on 2010-2014) | | | Computation 2 (Based on 2011-2015) | | | % Variation |
|------|-------|-------|-------|-------|-------|-------|-------|-------|
| | | Trend | Seasonal | Sum | Trend | Seasonal | Sum | |
| | | A | B | C = A + B | D | E | F = D + E | (F - C)/C *100 |
| 2011 | Jul | 7573 | 120 | 7693 | 7573 | 329 | 7902 | 2.72 |
| | Aug | 7397 | -102 | 7295 | 7397 | -97 | 7300 | 0.07 |
| | Sep | 7295 | 47 | 7342 | 7295 | 17 | 7312 | -0.41 |
| | Oct | 7161 | 319 | 7480 | 7161 | 19 | 7180 | -4.01 |
| | Nov | 7002 | 281 | 7283 | 7002 | 85 | 7087 | -2.69 |
| | Dec | 6849 | -27 | 6822 | 6849 | -33 | 6816 | -0.09 |
| 2012 | Jan | 6702 | 2 | 6704 | 6702 | 87 | 6789 | 1.27 |
| | Feb | 6591 | -294 | 6297 | 6591 | -23 | 6568 | 4.30 |
| | Mar | 6532 | -431 | 6101 | 6532 | -246 | 6286 | 3.03 |
| | Apr | 6523 | -49 | 6474 | 6523 | -67 | 6456 | -0.28 |
| | May | 6563 | -60 | 6503 | 6563 | -113 | 6450 | -0.82 |
| | Jun | 6657 | 193 | 6850 | 6657 | 40 | 6697 | -2.23 |
| | Jul | 6777 | 120 | 6897 | 6777 | 329 | 7106 | 3.03 |
| | Aug | 6823 | -102 | 6721 | 6823 | -97 | 6726 | 0.07 |
| | Sep | 6789 | 47 | 6836 | 6789 | 17 | 6806 | -0.44 |
| | Oct | 6728 | 319 | 7047 | 6728 | 19 | 6747 | -4.26 |
| | Nov | 6681 | 281 | 6962 | 6681 | 85 | 6766 | -2.82 |
| | Dec | 6644 | -27 | 6617 | 6644 | -33 | 6611 | -0.09 |
| 2013 | Jan | 6579 | 2 | 6581 | 6579 | 87 | 6666 | 1.29 |
| | Feb | 6483 | -294 | 6189 | 6483 | -23 | 6460 | 4.38 |
| | Mar | 6376 | -431 | 5945 | 6376 | -246 | 6130 | 3.11 |
| | Apr | 6266 | -49 | 6217 | 6266 | -67 | 6199 | -0.29 |
| | May | 6163 | -60 | 6103 | 6163 | -113 | 6050 | -0.87 |
| | Jun | 6071 | 193 | 6264 | 6071 | 40 | 6111 | -2.44 |
| | Jul | 5988 | 120 | 6108 | 5988 | 329 | 6317 | 3.42 |
| | Aug | 5938 | -102 | 5836 | 5938 | -97 | 5841 | 0.09 |
| | Sep | 5951 | 47 | 5998 | 5951 | 17 | 5968 | -0.50 |
| | Oct | 6040 | 319 | 6359 | 6040 | 19 | 6059 | -4.72 |
| | Nov | 6191 | 281 | 6472 | 6191 | 85 | 6276 | -3.03 |
| | Dec | 6448 | -27 | 6421 | 6448 | -33 | 6415 | -0.09 |
| 2014 | Jan | 6803 | 2 | 6805 | 6803 | 87 | 6890 | 1.25 |
| | Feb | 7191 | -294 | 6897 | 7191 | -23 | 7168 | 3.92 |
| | Mar | 7618 | -431 | 7187 | 7618 | -246 | 7372 | 2.57 |
| | Apr | 8046 | -49 | 7997 | 8046 | -67 | 7979 | -0.23 |
| | May | 8466 | -60 | 8406 | 8466 | -113 | 8353 | -0.63 |
| | Jun | 8884 | 193 | 9077 | 8884 | 40 | 8924 | -1.69 |

**Method VI:** The objective of this method is to gain an insight into the contribution of the trend and the seasonal components in the overall time series of the Consumer Durables and the Small

Cap sector. As we mentioned in this Section 5, this approach is based on comparison of the aggregate of the trend and the seasonal components of a time series over two different period of time. First, we construct a time series using the time series data for the period January 2010 to December 2014, and then we compute the trend and the seasonal components and their aggregate values. We refer to this computation as *Computation 1*. For the Consumer Durable sector time series, the trend, the seasonal and their aggregate values in *Computation 1* are noted in the columns *A*, *B* and *C* respectively in Table 15. Next, we construct the second time series using the time series data for the period January 2011 to December 2015 and repeat the computation of the trend, the seasonal and their aggregate values. We refer to this computation as *Computation 2*. For the Consumer Durables sector, the trend, the seasonal and their aggregate values in *Computation 2* are noted in the columns *D*, *E* and *F* respectively in Table 15. The percentages of variation of the aggregate values in both computations are noted for each month for the period July 2011 to June 2014. If there is a structural difference between the time series data in 2010 and 2015, then we expect that difference to be reflected in the aggregate of the trend and the seasonal values. The computations for the Small Cap sector are presented in Table 16.

**Observation:** From both Table 15 and Table 16, it is quite evident that the aggregate of the trend and the seasonal components had remained consistently the same over the period July 2011 to June 2014 for both the Consumer Durables and the Small Cap sector. This indicates that there have been no structural changes in the time series of these two sectors during the period January 2010 to December 2015. Since the change of the time series due to substitution of the 2010 data by 2015 data had virtually no impact on the trend and the seasonal components, we conclude that the impact of the random component is not significant, and the Consumer Durables and the Small Cap sectors time series is quite amenable for robust forecasting.

## 6. Related Work

Researchers have spent considerable effort in designing mechanisms for forecasting of daily stock prices. Applications of neural network based approaches have been proposed in many forecasting systems. Mostafa proposed neural network-based mechanism to predict stock market movements in Kuwait (Mostafa, 2010). Kimoto et al applied neural networks on historical accounting data and used various macroeconomic parameters for the purpose of prediction of variations in stock returns (Kimoto et al, 1990). Leigh et al proposed the use of linear regression and simple neural network models for forecasting the stock market indices in the New York Stock Exchange during the period 1981-1999 (Leigh et al, 2005). Hammad et al have demonstrated that artificial neural network (ANN) model can be trained to converge to an optimal solution while it maintains a very high level of precision in forecasting of stock prices (Hammad et al, 2009). Dutta et al demonstrate the application of ANN models for forecasting Bombay Stock Exchange's SENSEX weekly closing values for the period of January 2002- December 2003 (Dutta et al, 2006). Tsai and Wang found observations that highlighted the fact that Bayesian Network-based approaches have better forecasting power than traditional regression and neural network-based approaches (Tsai & Wang, 2009). Tseng et al deployed traditional time series decomposition (TSD), HoltWinters (H/W) models, Box-Jenkins (B/J) methodology and neural network- based approach on 50 randomly chosen stocks during September 1, 1998 - December 31, 2010 for forecasting the future values of the stock prices (Tseng et al, 2012). It has been observed that forecasting errors are lower for B/J, H/W and normalized neural network model, while the errors are appreciably larger for time series decomposition and non-normalized neural network models. Moshiri and Cameron presented a Back Propagation Network (BPN) with econometric models to forecast inflation using (i) Box-Jenkins Autoregressive Integrated Moving Average (ARIMA) model, (ii) Vector Autoregressive (VAR) model and (iii) Bayesian Vector Autoregressive (BVAR) model (Moshiri, & Cameron, 2010). Phua et al deployed ANNs with genetic algorithms for the purpose of predicting the stock prices in Singapore Stock Exchange (Phua et al, 2000). The result was promising with a forecast accuracy of 81% on the average. Hutichinson et al proposed a non-parametric method for estimating the pricing formula of a derivative that applied the principles of *learning networks* (Hutchinson et al, 1994). The inputs to the network were the primary economic variables that influence the derivative price, e.g., the current fundamental asset price, the strike price, the time to maturity etc. The derivative price was defined to be the output into which the learning network maps the inputs. The data used were the daily closing prices of S&P 500 futures and the options for the 5-year period from January 1987 to December 1991. The authors have compared their results with the parametric derivative pricing formula and the found the results quite promising. Thenmozhi examined the nonlinear nature of the Bombay Stock Exchange time series using chaos theory (Thenmozhi, 2001). The study examined the Sensex returns time series from August 1980 to September 1997 and showed that the daily returns and weekly returns of the BSE sensex are characterized by nonlinearity and the time series is weakly chaotic.

ANN and Hybrid systems are particularly effective in forecasting stock prices for stock time series data. A large number of work have been done based on ANN techniques for stock market prediction (Shen et al, 2007; Jaruszewicz & Mandziuk, 2004; Ning et al, 2009; Pan et al, 2005; Hamid & Iqbal, 2004; Chen, et al, 2005; Chen et al, 2003; Hanias et al, 2007; de Faria, 2009). Many applications of hybrid systems in stock market time series data analysis have also been proposed in the literature (Wu et al, 2008; Wang & Nie, 2008; Perez-Rodriguez et al, 2005; Leung et al, 2000; Kim, 2004).

In contrast to the work mentioned above, our approach in this paper is based on structural decomposition of a time series to study the behavior of two different sectors of the Indian economy – the Consumer Durables sector and the Small Cap sector. By decomposition of the time series of these two sectors for the period January 2010 – December 2015, we have demonstrated the fundamental differences between them. We found that while the seasonal component is much stronger in the Consumer Durables sector time series, the time series of the Small Cap sector had a dominant random component. Besides illustrating the fundamental differences between these time series, we have proposed five robust forecasting techniques and a quantitative framework for analyzing any change in behavior of the constituents of a time series over a long period of time so as to have an idea how effectively the time series future values may be predicted. We have computed the relative accuracies of each of the forecasting techniques, and also have critically analyzed under what situations a particular technique performs better than the other techniques. Our proposed framework of analysis can be used as a broad approach for forecasting the behavior of other stock market indices in India.

## 7. Conclusion

In this paper, we proposed a time series decomposition-based approach for deeper understanding and analysis of two sectors of the Indian economy – the Consumer Durables sector and the Small Cap sector. We have demonstrated that decomposition results provide us insights about the fundamental characteristics of the sectors which in turn can enable the investors in making wise and efficient investment decisions about their portfolios. Using our proposed decomposition approach, we have also validated two hypotheses – (i) the Consumer durables sector has a strong seasonal component and (ii) the Small Cap sector is characterized by the presence of a strong random component. After analyzing the time series decomposition results and validating the hypotheses, we have proposed five robust forecasting techniques and a quantitative framework for analyzing the behavior of the structural constituents of a time series. We have presented detailed results on the performance of each of the forecasting methods and also critically analyzed why certain method has performed best compared to the others, for what type of time series and under what situations. The proposed structural decomposition and analysis approach provided enough insights about the way the constituents of the time series for the two sectors had behaved over the period under investigation, i.e., January 2010 – December 2015. It has been demonstrated clearly that the time series of both the sectors are quite amenable for robust and accurate forecasting even in presence of a dominant random component.

The results obtained from the above analysis is extremely useful for portfolio construction. When we perform this analysis for other sectors as well, it will help portfolio managers and individual investors to identify which sector, and in turn which stock, to buy/sell in which period. It will also help in identifying which sector, and hence which stock, is dominated by the random component and thus is speculative in nature.

## References

Chen, A.-S., Leung, M. T. & Daouk, H. (2003). Application of neural networks to an emerging financial market: forecasting and trading the Taiwan stock index. *Operations Research in Emerging Economics*, 30(6), 901– 923.
Chen, Y., Dong, X. & Zhao, Y. (2005). Stock index modeling using EDA based local linear wavelet neural network. *Proceedings of International Conference on Neural Networks and Brain*, 1646–1650.
Coghlan, A. (2015). *A Little Book of R for Time Series*, Release 02.
   Available online: https://media.readthedocs.org/pdf/a-little-book-of-r-for-time-series/latest/a-little-book-of-r-for-time-series.pdf. (last accessed on: May 9, 2016).

Data Chaudhuri, T., Ghosh, I., & Eram, S. (2016). Predicting stock returns of mid cap firms in India – an application of random forest and dynamic evolving neural fuzzy inference system. *Proceedings of the 2$^{nd}$ National Conference on Advances in Business Research and Management Practices (ABRMP'16)*, Kolkata, India, January 8 -9, 2016.
Available online: http://papers.ssrn.com/sol3/papers.cfm?abstract_id=2709913 (last accessed on: May 9, 2016)

de Faria, E., Albuquerque, M. P., Gonzalez, J., Cavalcante, J. & Albuquerque, M. P. (2009). Predicting the Brazilian stock market through neural networks and adaptive exponential smoothing methods. *Expert Systems with Applications*.

Dutta, G. Jha, P., Laha, A. & Mohan, N. (2006). Artificial neural network models for forecasting stock price index in the Bombay stock exchange. *Journal of Emerging Market Finance*, 5, 283-295.

Hamid, S. A., & Iqbal, Z. (2004). Using neural networks for forecasting volatility of S&P 500 index futures prices. *Journal of Business Research*, 57(10), 1116–1125.

Hammad, A., Ali, S. & Hall, E. (2009). Forecasting the Jordanian stock price using artificial neural network. Available online:
http://www.min.uc.edu/robotics/papers/paper2007/Final%20ANNIE%2007%20Souma%20Alhaj%20Ali%206p.pdf). (last accessed on: May 9, 2016).

Hanias, M., Curtis, P. & Thalassinos, J. (2007). Prediction with neural networks: the Athens stock exchange price indicator. *European Journal of Economics, Finance and Administrative Sciences*, 9, 21–27.

Hutchinson, J. M., Lo, A. W., & Poggio, T. (1994). A nonparametric approach to pricing and hedging derivative securities via learning networks. *Journal of Finance*, 49(3), 851-889.

Ihaka, R. & Gentleman, R. (1996). A language for data analysis and graphics. Journal of Computational and Graphical Statistics, 5(3), 299-314.

Jaruszewicz, M. & Mandziuk, J. (2004). One day prediction of Nikkei index considering information from other stock markets. *Proceedings of the International Conference on Artificial Intelligence and Soft Computing*, 3070, 1130–1135.

Kimoto, T., Asakawa, K., Yoda, M. & Takeoka, M. (1990). Stock market prediction system with modular neural networks. *Proceedings of the IEEE International Conference on Neural Networks*, 1-16.

Leigh, W., Hightower, R. & Modani, N. (2005). Forecasting the New York stock exchange composite index with past price and interest rate on condition of volume spike. *Expert Systems with Applications*, 28, 1-8.

Leung, M. T., Daouk, H. & Chen, A.-S. (2000). Forecasting stock indices: a comparison of classification and level estimation models. *International Journal of Forecasting*, 16(2), 173–190.

Moshiri, S. & Cameron, N. (2010). Neural network versus econometric models in forecasting inflation. *Journal of Forecasting*, 19, 201-217.

Mostafa, M. (2010). Forecasting stock exchange movements using neural networks: empirical evidence from Kuwait. *Expert Systems with Application*, 37, 6302-6309.

Ning, B., Wu, J., Peng, H. & Zhao, J. (2009). Using chaotic neural network to forecast stock index. *Advances in Neural* Networks, 5551, 870–876.

Pan, H., Tilakaratne, C. & Yearwood, J. (2005). Predicting the Australian stock market index using neural networks exploiting dynamical swings and intermarket influences. *Journal of Research and Practice in Information Technology*, 37(1), 43–55.

Perez-Rodriguez, J. V., Torra, S. & Andrada-Felix, J. (2005). Star and ANN models: forecasting performance on the Spanish IBEX-35 stock index. *Journal of Empirical Finance,* 12(3), 490–509.

Phua, P. K. H., Ming, D., & Lin, W. (2000). Neural network with genetic algorithms for stocks prediction. 5th Conference of the Association of Asian-Pacific Operations Research Societies, Singapore.

Sen J. and Datta Chaudhuri, T. (2016 a). "Decomposition of Time Series Data of Stock Markets and its Implications for Prediction – An Application for the Indian Auto Sector", *Proceedings of the 2nd National Conference on Advances in Business Research and Management Practices (ABRMP'16)*, Kolkata, India, January 8 -9, 2016.
Available online at: https://arxiv.org/abs/1601.02407 (last accessed on: May 9, 2016)

Sen, J. and Datta Chaudhuri, T. (2016b). A framework for predictive analysis of stock market indices – a study of the Indian auto sector. *Calcutta Business School (CBS) Journal of Management Practices*, 2(2), 1 – 19.
Available online: https://arxiv.org/abs/1604.04044 (last accessed on May 9, 2016)

Shen, J., Fan, H. & Chang, S. (2007). Stock index prediction based on adaptive training and pruning algorithm. *Advances in Neural Networks*, 4492, 457–464.

Thenmozhi, M. (2006). Forecasting stock index numbers using neural networks. Delhi Business Review, 7(2), 59-69.

Tsai, C.-F. & Wang, S.-P. (2009). Stock price forecasting by hybrid machine learning techniques. *Proceedings of International Multi Conference of Engineers and Computer Scientists*, 1.

Tseng, K-C., Kwon, O., & Tjung, L. C. (2012). Time series and neural network forecast of daily stock prices. *Investment Management and Financial Innovations*, 9(1), 32-54.

Wang, W. & Nie, S. (2008). The performance of several combining forecasts for stock index. *International Workshop on Future Information Technology and Management Engineering*, 450– 455.

Wu, Q., Chen, Y. & Liu, Z. (2008). Ensemble model of intelligent paradigms for stock market forecasting. *Proceedings of the IEEE 1st International Workshop on Knowledge Discovery and Data Mining*, 205 – 208, Washington, DC, USA.

Zhu, X., Wang, H., Xu, L. & Li, H. (2008). Predicting stock index increments by neural networks: the role of trading volume under different horizons. *Expert Systems Applications*, 34(4), 3043–3054.